\documentclass[aps,prl,superscriptaddress,twocolumn,showpacs,floatfix]{revtex4}

\usepackage{epsfig}
\usepackage{bm}
\usepackage{graphicx}
\usepackage{amssymb}
\usepackage{dcolumn}

\begin{document}

\title{Anisotropy-induced ordering in the quantum $J_1-J_2$ 
antiferromagnet}

\author{Tommaso Roscilde}
\affiliation{Department of Physics and Astronomy, University of Southern
California, Los Angeles, CA 90089-0484}
\author{Adrian Feiguin}
\affiliation{Department of Physics and Astronomy, University of California,
Irvine, CA 92697}
\author{Alexander L. Chernyshev}
\affiliation{Department of Physics and Astronomy, University of California,
Irvine, CA 92697}
\author{Shiu Liu}
\affiliation{Department of Physics and Astronomy, University of California,
Irvine, CA 92697}
\author{Stephan Haas}
\affiliation{Department of Physics and Astronomy, University of Southern
California, Los Angeles, CA 90089-0484}
\date{\today}

\begin{abstract}
We study the effect of spin anisotropies on  
a frustrated quantum antiferromagnet 
using the $J_1$-$J_2^{XXZ}$ model on the square lattice.  
The $T=0$ and finite-$T$  phase diagrams of this model are obtained utilizing
spin-wave theory, exact diagonalization, and quantum Monte Carlo.
We find that anisotropic frustration tends 
to stabilize $XY$- and Ising-like ordered phases, 
while the disordered spin-liquid phase is restricted to a small 
region of the phase diagram.
The ordered phases are separated by first-order transitions and exhibit
a non-trivial reentrance phenomenon.
\end{abstract}


\maketitle

Frustrated quantum magnets attract significant interest 
because of the spin-liquid and novel ordered phases they may
exhibit.\cite{MisguichL03,Kawamura98,CCL90}  
The macroscopic degeneracy of the ground state in such magnets
makes them very sensitive to additional interactions that
may lead to various unconventional ordered states.
Perhaps the most extensively investigated, yet highly
controversial model is the $S=1/2$ $J_1$-$J_2$ Heisenberg model 
on the square lattice with competing
nearest-neighbor, $J_1$, and next-nearest-neighbor, $J_2$,
interactions.\cite{isotropic} 
The recent synthesis of compounds that 
can be closely described by
the  two-dimensional (2D) $J_1$-$J_2$ Hamiltonian\cite{Melzietal01} 
has also fueled
interest in the properties of this basic model of magnetic frustration.
In that respect, the presence of spin anisotropies in real systems
raises the question of how robust the behavior of the 
isotropic $J_1$-$J_2$ model is against such perturbations. 

In the present paper we study the effect of spin anisotropies 
on the properties of frustrated quantum antiferromagnets 
using a generalization of the $J_1$-$J_2$ model,
in which the frustrating next-nearest-neighbor interaction
is anisotropic. Such a $J_1$-$J_2^{XXZ}$ model is given by:
\begin{equation}
 \hat{\cal H} = \frac{J_1}{2}\sum_{n.n.}
 {\bm S}_i\cdot{\bm S}_j+ 
 \frac12\sum_{n.n.n.} \bigg(J_2^z S_i^z S_k^z+J_2^\perp
 S_i^+ S_k^-\bigg) \ , 
 \label{e.j1j2xxz}
 \end{equation} 
\noindent
where ${\bm S}_{i}$ is the spin operator, 
the sites $i, j, k$ are on the square lattice, and the summation runs
over all nearest-neighbor ($n.n.$) or next-nearest-neighbor ($n.n.n.$)
sites. We hereafter use the dimensionless ratios
$\alpha_z =  J_2^{z}/J_1$ and 
$\alpha_{\perp} =  J_2^{\perp}/J_1$.

For the extensively studied isotropic case $\alpha_{\perp} =\alpha_z$,
analytical and numerical works have suggested 
the existence of a $T=0$, 
non-magnetic gapped phase for $0.4\alt J_2/J_1\alt 0.6$, separating
the N\'eel state from the collinear state.\cite{isotropic} 
The absence of long-range order at any finite temperature  in
2D systems for the isotropic $J_1$-$J_2$ case is dictated
by the Mermin-Wagner theorem. 
In the collinear phase, however, the 2D $J_1$-$J_2$ model  
has been predicted to exhibit an Ising-like Chandra-Coleman-Larkin (CCL) 
transition \cite{CCL90} at $T>0$  with spontaneous breaking 
of the discrete lattice rotational symmetry. This unusual scenario 
has recently received support from numerical and analytical 
studies.\cite{Capriottietal03}

\begin{figure}[t] 
\includegraphics[angle=0,width=7cm]{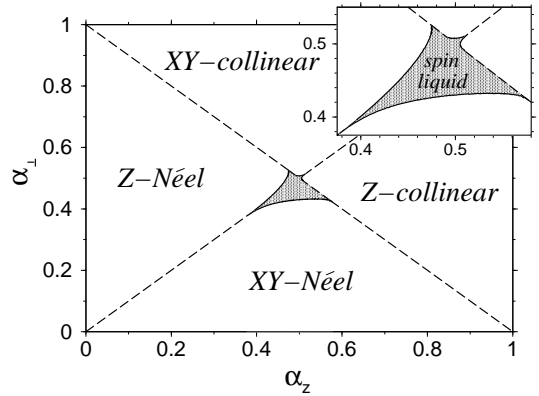}
 \caption{\label{clphdiagr} Zero-temperature phase diagram of the $S=1/2$
$J_1$-$J_2^{XXZ}$ model  within linear spin-wave theory. 
Solid and dashed lines denote the second- and first-order transitions,
respectively. Inset: non-magnetic region magnified.}
\null \vspace{-.4cm}
\end{figure}
In this Letter we show that even a small anisotropy
in the frustrating ($n.n.n.$) coupling leads to strong 
deviations from the behavior of the isotropic 
$J_1$-$J_2$ model. The main effect
of anisotropy is to energetically favor 
the sector of spin space in which the frustration is minimal.
This tendency of spins to avoid frustration leads to somewhat
counter-intuitive results, as, {\it e.g.}, Ising-like anisotropy 
in $J_2$ favoring the $XY$ phase and vice versa.
In the case of the unfrustrated 2D
Heisenberg antiferromagnet, an arbitrarily small anisotropy 
is found to stabilize finite-temperature ordered phases. 
\cite{Cuccolietal03}
In this work we show that the anisotropically frustrated
$J_1$-$J_2^{XXZ}$ model is generally 
more ordered than its isotropic counterpart.

The $T=0$ phase diagram of classical 
($S=\infty$) $J_1$-$J_2^{XXZ}$ model is shown in
Fig. \ref{clphdiagr}.   
We find four phases characterized by different order parameters,
$m^{\alpha}_{\bm q} = \left\langle \sum_{\bm r}
 e^{i \bm q \bm r} S_{\bm r}^\alpha \right\rangle$, hereafter indicated as: 
{\it (i)} {\it XY-N\'eel} phase with  $m^{xy}_{(\pi,\pi)}$,
{\it (ii)} {\it Z-collinear} phase with  $m^{z}_{(\pi,0)}$ or   
$m^{z}_{(0,\pi)}$, {\it (iii)} {\it XY-collinear} phase with  
$m^{xy}_{(\pi,0)}$ or $m^{xy}_{(0,\pi)}$, and 
{\it (iv)} {\it Z-N\'eel} phase
with $m^{z}_{(\pi,\pi)}$. The diagonal line 
$\alpha_{\perp} = \alpha_{z}$  
corresponds to the isotropic $J_1$-$J_2$ model.
All four phases are separated by first-order 
transition lines.
Taking into account quantum corrections within linear spin-wave 
theory results in the opening of a {\it spin-liquid
region} where all the order parameters vanish.
Along the main diagonal this reproduces the well-known
results for the isotropic case.\cite{isotropic}  
Our analysis suggests that in the vicinity of the $\alpha_\perp
=\alpha_z$ line the $T=0$ transitions from the
ordered phases to the spin-liquid phase is of second order. 
However, we also find that the collinear phases are separated from
the spin-liquid phase mostly by the {\it first-order} transition
lines, as shown in Fig. \ref{clphdiagr}. This means that the $XY$- and
{\it Z-collinear} orders are particularly robust, so that 
even quantum-renormalized order parameters vanish with a finite 
jump when crossing such boundaries. 
 
Given the anisotropic nature of the ordered phases found in the
$T=0$ phase diagram, one can anticipate qualitatively the finite
temperature behavior.
In the Ising regions {\it magnetic} order is expected to survive 
at $T>0$   up  to a second-order transition point. 
{\it Topological} order is instead expected at $T>0$
in the $XY$ regions up to a Berezinskii-Kosterlitz-Thouless (BKT) 
transition. 

The study of the quantum $J_1$-$J_2^{XXZ}$ 
model at finite temperature by quantum Monte Carlo (QMC) techniques
is generally precluded by the sign problem \cite{HeneliusS00} 
due to the transverse frustrating  coupling $\alpha_{\perp}$. 
However, by removing that term completely
we obtain the sign-problem-free  
$J_1$-$J_2^{Z}$ model, where the frustration is Ising-like,
corresponding to the $\alpha_{\perp}=0$ 
line of the phase diagram in Fig. \ref{clphdiagr}.
A comprehensive study of this  limiting case is, therefore,
very important since its thermodynamics can shed 
light on the main features of a large region of the phase diagram
($\alpha_{\perp} < \alpha_{z}$).
Moreover, in this limit of $J_2^{z}$-only frustration, the 
Hamiltonian of Eq. (\ref{e.j1j2xxz}) is also relevant for the 
broad class of strongly correlated {\it bosonic} systems. 
In fact, this $J_1$-$J_2^{Z}$ model can be 
mapped exactly onto the frustrated hard-core Bose-Hubbard
model.\cite{Hebert01}  

\begin{figure}[t] 
\includegraphics[bbllx=0pt,bblly=40pt,bburx=516pt,bbury=450pt,
angle=0,width=8cm]{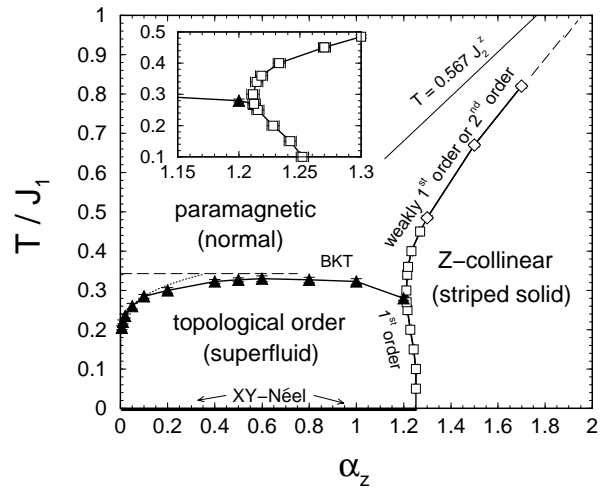}
 \caption{\label{qphdiagr} Phase diagram of the $S=1/2$
$J_1$-$J_2^{Z}$ model. Magnetic (and corresponding bosonic) 
phases are indicated.
The horizontal dashed line marks the BKT critical temperature
of the $S=1/2$ XY model \cite{HaradaK98}, the dotted line is a
logarithmic fit for small $\alpha_z$ (see text). 
Inset: zoom on the tricritical point.}
\null \vspace{-.4cm}
\end{figure} 
We have thus investigated the thermodynamics of the 
$S=1/2$ $J_1$-$J_2^{Z}$ model by means of the Stochastic Series Expansion 
QMC based on the directed-loop
algorithm.\cite{SyljuasenS02} A large interval
of frustration values, $0 < \alpha_z < 1.7$, has been scanned using 
different lattices $L\times L$  
up to $L=96$ to perform an extensive finite-size 
scaling analysis.
The resulting phase diagram of the $J_1$-$J_2^{Z}$ model 
is shown in Fig. \ref{qphdiagr}. 
As main features we observe the occurrence of the topologically
ordered phase (quasi-long-range {\it XY}-N\'eel) 
for small $\alpha_z$ and the {\it
  Z}-collinear  order for  large $\alpha_z$.
These two phases are separated by a 
first-order transition line.
Phases with finite-temperature topological
order and {\it Z}-collinear order
in the spin language correspond, in the bosonic language,
to quasi-long-range superfluidity and
striped solid order, respectively. 
 
The study of the first-order transition line has been 
performed by making use of a quantum parallel tempering 
technique\cite{Senguptaetal02}
to overcome critical slowing down at first-order
transition points. \cite{Janke98} 
In the collinear phase 
we have also used  thermal parallel tempering 
\cite{Marinari96} to overcome the well-known 
loss of efficiency of standard cluster algorithms
in Ising-like frustrated antiferromagnets. 
\cite{Kandeletal90}

A remarkable feature of the phase diagram is that even an infinitesimal
frustration, $\alpha_z \ll 1$, induces a finite-temperature BKT transition. 
The critical temperature $T_{\rm BKT}$
has been extracted through scaling analysis of the helicity
modulus $\Upsilon$ and of the transverse structure factor
$S^{xx(yy)}(\pi,\pi)$. \cite{HaradaK98,Cuccolietal03} As expected
for weakly anisotropic systems, \cite{Cuccolietal03}
$T_{\rm BKT}$ depends logarithmically on the frustration as
$T_{\rm BKT} \approx 4\pi\rho_s/(C + |\ln\alpha_z|)$,
where $\rho_s$ is the spin stiffness of the unfrustrated Heisenberg
antiferromagnet and $C$ is a constant. A fit of our data 
to the above law for $\alpha_z \leq 0.1$ 
gives $\rho_s = 0.175(2)$ in excellent agreement with the 
best available estimates, \cite{Sandvik97} and $C = 5.4(1)$. 
For larger frustration, the BKT transition temperature
reaches a maximum which falls very close to the 
critical temperature of the $S=1/2$ XY model,
$T_{\rm BKT}/J = 0.3427(2)$, \cite{HaradaK98} indicating that
the $J_1$-$J_2^{Z}$ model with frustration 
$\alpha_z \approx 0.65$ is an almost ideal realization
of the quantum XY model. 

\begin{figure}[t] 
\includegraphics[angle=0,width=8cm]{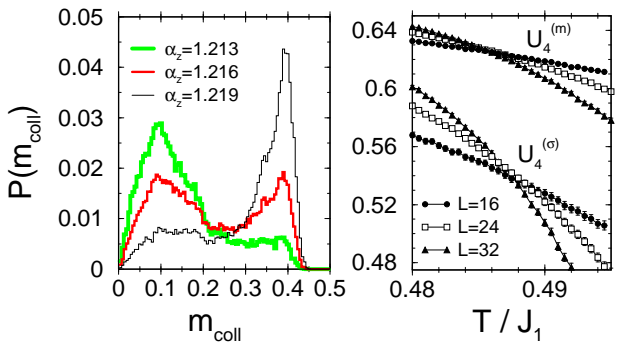}
 \caption{\label{f.firstorder} Left panel: the double-peak structure
in the histograms of the collinear
magnetization $m_{\rm coll}$ 
for system size $L=32$ and temperature $T/J_1=0.36$. 
Right panel: Binder's fourth cumulant of the 
 collinear magnetization and the CCL order parameter for $\alpha_z = 1.3$.}
\null \vspace{-.9cm}
\end{figure} 
At zero temperature, the $XY$-N\'eel and the
$Z$-collinear phases are separated by a
first-order transition at the critical value 
$\alpha_z^{(c)} = 1.252(5)$. We notice that 
quantum fluctuations cause a 25\% shift of this 
critical value with respect to the classical value
$\alpha_z^{(c)} = 1$, thus promoting the N\'eel
phase against the collinear phase. 
When increasing the temperature the first order transition line
shows a {\it reentrant behavior}, as seen in the inset of 
Fig. \ref{qphdiagr}. For $1.21 \alt \alpha_z \alt 1.25$, 
when temperature is increased, the topological order becomes unstable
to the onset of the collinear magnetic order, which 
carries a higher entropy content. One can argue that
the bending of the first-order transition line reflects the 
reduced role of quantum fluctuations, responsible for shifting 
the critical value $\alpha_z^{(c)}$, as the temperature is 
raised above zero. 
We notice that similar phase reentrance phenomena have recently been 
observed in related magnetic and bosonic systems. \cite{Schmid}

The first-order transition line meets the BKT line
at the point $(\alpha_{z},T/J_1) = (1.210(5),0.28(1))$,
above which it maintains its first-order nature
up to a temperature $T/J_1 \approx 0.36$.
This is shown in Fig. \ref{f.firstorder}, where the
phase coexistence manifests itself in the double-peak
structure of the distribution for the collinear 
magnetization $m_{\rm coll} = m^z_{(\pi,0)} + m^z_{(0,\pi)}$ 
around the transition point. We have checked that the 
double-peak feature persists when increasing the 
lattice size up to $L=48$.
For $T \gtrsim 0.36 J_1$, determining the 
order of the transition to collinear order becomes more complicated. 
In fact the distribution of the order parameter 
$m_{\rm coll}$ loses the strong two-peak feature, 
suggesting the transition to be {\it weakly first-order}. 
For $\alpha_z \gg 1$ the model reduces
to two disconnected Ising models,
living on the two sublattices of the original square lattice. 
In this limit, according to Onsager's solution,  we 
should expect a 2D Ising transition
at a critical temperature $T_{\rm coll} = 
2 J_2^{z} S^2 /\ln(1+\sqrt{2}) = 0.567 ~J_2^{z}$.
The crossover from a weakly first-order to a second-order 2D Ising 
transition might happen with a
continuous change of the critical exponents
for increasing $\alpha_z$, as argued to occur in the 
frustrated 2D Ising antiferromagnet. \cite{BinderL80}
 
We have determined the transition points $T_{\rm coll}$ for  
$T \gtrsim 0.36$ by the crossing of Binder's fourth
cumulant $U_4^{(m)} = 1 - \langle m_{\rm coll}^4 \rangle/ 
3 \langle m_{\rm coll}^2\rangle^2$.
In Fig. \ref{f.firstorder} we observe 
that, for $\alpha_z = 1.3$ as well as for the other 
cases studied, 
the values of $U_4^{(m)}(L)$ for different lattice sizes
do not cross at the 2D Ising critical value 
$U_{4,c}= 0.6107$ \cite{KamieniarzB93}, which
suggests the universality class of the transition
to be distinct from 2D Ising.
  
An interesting question is whether the CCL transition
\cite{CCL90} of the isotropic $J_1$-$J_2$ 
antiferromagnet, with spontaneous breaking 
of the discrete lattice rotational symmetry,
survives in the collinear
phase of the anisotropic $J_1$-$J_2^Z$ model.
The order parameter for the transition
reads:\cite{CCL90,Capriottietal03}
$\sigma = \langle ({S}^z_{i,j}-{ S}^z_{i+1,j+1})
({ S}^z_{i+1,j}-{ S}^z_{i,j+1}) \rangle/4$.
We have studied the scaling of Binder's fourth
cumulant related to the order parameter $\sigma$, 
$U_4^{(\sigma)} = 1 - \langle \sigma_{\rm coll}^4 \rangle/ 
3 \langle \sigma_{\rm coll}^2\rangle^2$.  
As shown by the crossing of the Binder's cumulant 
in Fig. \ref{f.firstorder}, the CCL transition
seems to occur at the same (or slightly
higher) temperature with respect to the magnetic transition. 
\begin{figure}[t] 
\includegraphics[angle=0,width=8cm]{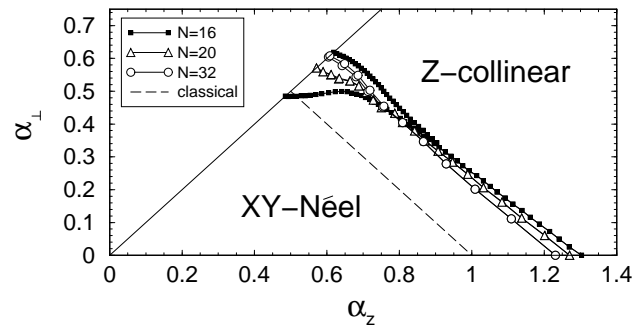}
 \caption{\label{GSphdiagr} Zero-temperature phase diagram of the
  $J_1$-$J_2^{XXZ}$ model from exact diagonalization on clusters
  of size $N$.}
\null \vspace{-.9cm}
\end{figure} 

Let us now return to the general $J_1$-$J_2^{XXZ}$ model to
investigate how the ground state 
properties evolve in presence of the transverse component of
frustration, $\alpha_{\perp}\leq \alpha_{z}$. 
Making use of exact diagonalization on lattices of 
$N=$16, 20, and 32 spins we have determined the dependence of the order 
parameters, $m^{x}_{(\pi,\pi)}$ and $m_{\rm coll}$, on  
$\alpha_z$ at fixed $\alpha_{\perp}$. Using a dense mesh
of points, we obtain a qualitative 
finite-size estimate of the phase
boundaries from the vanishing of the second derivatives
$\partial^2 m^{\alpha}_{(\bm q)} / \partial \alpha_z^2$,
marking an inflection point in the order parameters.
In this way we have obtained the $T=0$ phase diagram
shown in Fig. \ref{GSphdiagr}. For sufficiently low transverse
frustration ($\alpha_{\perp}\lesssim 0.4$) 
both order parameters display an inflection at the 
same point,
suggesting that the transverse frustration does not 
introduce any intermediate phase. The sharpness
of the peaks in the derivatives of the 
order parameters suggests that the
transition maintains the first-order nature found in
the $J_1$-$J_2^{z}$ model. This first-order
transition line runs roughly parallel to the classical line,
being shifted to higher $\alpha_z$ by about 20\% due to 
quantum fluctuations. As the transverse
frustration is increased, the inflection points of the 
order parameters separate from each other, 
leaving space for an intermediate region with no
magnetic order, continuously connected with the 
spin-liquid phase in the isotropic limit. 
Nonetheless, we observe the intermediate non-magnetic
phase to shrink dramatically as the cluster size is increased.
The small size of the clusters does not
allow us to conclude  on the real extension of the 
intermediate non-magnetic phase and on the nature of the phase 
transition line(s) for $\alpha_{\perp} \gtrsim 0.4$. Further
investigations of this issue are currently in progress. 
A suggestive scenario is that
the transition line between the collinear and the non-magnetic
phase remains of first order all the way to the isotropic 
limit, in agreement with the predictions of previous
works \cite{Chubukov91}. 
  
\begin{figure}[t] 
\includegraphics[bbllx=100pt,bblly=30pt,bburx=516pt,bbury=500pt,
angle=0,width=5.5cm]{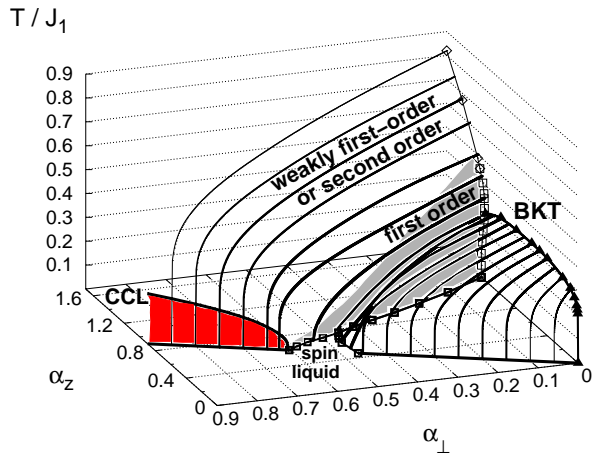}
 \caption{\label{3Dphdiagr} Global
 phase diagram of the $J_1$-$J_2^{XXZ}$ model. The leftmost 
 filled area represents the region with broken rotational
 symmetry in lattice space but unbroken symmetry 
 in spin space; data for the CCL transition are from
 Ref. \onlinecite{Capriottietal03}.}
\end{figure}
We finally propose a global phase diagram
for the $J_1$-$J_2^{XXZ}$ model, shown in 
Fig. \ref{3Dphdiagr}, by merging the 
the $T=0$ data with the finite-temperature data.
Here the finite-temperature $Z$-collinear and topological phases
are separated by a first-order transition 
surface, and upper-bounded by a BKT transition surface and a
weakly-first-order/second-order surface, respectively.
For $\alpha_z =\alpha_{\perp}$ the above critical surfaces
vanish. Moreover, in the high-frustration region 
of $\alpha_{\perp} \approx \alpha_z \approx 0.5$
we adopt the picture of an extended spin-liquid region
where all transition temperatures are also vanishing. 
For $\alpha_{\perp}\to\alpha_z$ outside 
the spin-liquid region, the critical 
temperatures $T_c$ vanish with a very 
steep slope, following a logarithmic dependence on the 
anisotropy  $\Delta = |\alpha_z-\alpha_{\perp}| \ll 1$, 
$T_{c} \sim |\ln(\Delta)|^{-1}$.
This dependence is obtained through the mean-field 
condition \cite{CHN88,Cuccolietal03} $k_B T_{c} \sim
J_1 \Delta [\xi^{(J_1-J_2)}]^2$, assuming
an exponential divergence of the correlation length of the 
isotropic $J_1$-$J_2$ model,
$\xi^{(J_1-J_2)} \approx T^{-1} \exp(2\pi\rho_s^{(J_1-J_2)}/T)$ 
(as found, {\it e.g.}, by modified spin-wave 
theory \cite{IvanovIvanov92}). At the spin-liquid
region boundaries, instead, the spin stiffness $\rho_s$ 
vanishes; to account for that, along the boundary between
the spin-liquid and the N\'eel region we introduce a 
$\Delta$-dependent stiffness in the mean-field equation, 
$\rho_s \sim (\Delta - \Delta_c)^{\beta}$, 
where $\Delta =\Delta_c$ marks the boundaries of the 
spin-liquid region. In drawing the phase diagram we
have taken $\beta=1$, as suggested by the data in 
Ref. \onlinecite{EinarssonS95}. In this way we get 
a much smoother vanishing of $T_c$, $T_{c} \sim 
(\Delta - \Delta_c)/|\ln(\Delta-\Delta_c)]|$.  
For the isotropic case $\alpha_z = \alpha_{\perp}$ 
and in the collinear phase, 
the only finite-$T$ transition line is the CCL one, 
recently estimated for $S=1/2$
in Ref. \onlinecite{Capriottietal03}. Having observed
the CCL transition to occur very close to the 
magnetic transition for $\alpha_{\perp} = 0$, 
we expect the CCL transition surface to quickly merge with the 
steep magnetic transition surface as
$\alpha_{\perp} \lesssim \alpha_{z}$. 
Thus, we suggest that the CCL transition 
exists as a distinct thermodynamic 
feature only in a very close vicinity of the 
isotropic limit.
 
In summary, we have shown that the Heisenberg antiferromagnet on the
square lattice with anisotropic next-nearest-neighbor frustration 
shows a very rich phase diagram characterized by several
finite-temperature transitions, which we studied using QMC, exact
diagonalization, and spin-wave theory.
The competition between ordered phases leads to non-trivial
reentrance phenomena, which is also common to related bosonic
systems. 
Since anisotropic spin-spin interactions are a general 
feature of real magnets, the ordering effect due to anisotropic 
frustration is a realistic mechanism to explain magnetic
transitions in recently synthetized frustrated quantum 
antiferromagnets.

Fruitful discussions with L. Spanu, S. White, and M. Zhitomirsky are 
gratefully acknowledged. We acknowledge support
by DOE under grant  DE-FG03-01ER45908 (T.R. and S.H.), by NSF under
grant DMR-0311843 (A.F. and S.L.), 
and by the ACS Petroleum Research Fund (A.L.C.).

\end{document}